# Optical control and decoherence of spin qubits in quantum dots

Paweł Machnikowski[a], Anna Grodecka[a,b], Carsten Weber[b] and Andreas Knorr[b]

[a]Institute of Physics, Wrocław University of Technology, 50-370 Wrocław, Poland
[b]Institut für Theoretische Physik, Technische Universität Berlin, 10623 Berlin, Germany

E-mail: Pawel.Machnikowski@pwr.wroc.pl

**Abstract**
We discuss various methods of all-optical spin control in semiconductor quantum dots. We present different ways of rotating a single confined electron spin by optical coupling to a trion state. We also discuss a method for controlling the polarization of a confined exciton via a two-photon transition. Finally, we analyze the effect of phonon-induced decoherence on the fidelity of these optical spin control protocols.

## 1. Introduction

Single electron spins confined in semiconductor quantum dots (QDs) seem to be promising candidates for implementing quantum information processing schemes. The main advantage of these systems is their relatively long coherence time. Many control protocols for spins in QDs have been developed, mostly based on applying external voltages and exploiting exchange interactions.

It is, however, also possible to control spins by optical means. In quantum dots this can be done by inducing charge dynamics dependent on the spin state of an electron, which is possible due to the optical selection rules and Pauli exclusion principle. Such control methods lead to very short switching times, even on picosecond time scales. A few control protocols of this kind have been proposed [1-6]. Their common feature is the optical coupling of both spin states to a trion (charged exciton) state. This opens a quantum pathway between the two spin states via a Raman transition. In this way optical coherent control of a spin in a QD becomes possible.

In this paper, these optical spin control schemes will be reviewed, mostly from a theoretical point of view. Schemes for rotating electron spins will be discussed, as well as methods for optical control of exciton polarization (by coupling to the biexciton state). This review will be followed by an analysis of decoherence processes accompanying the optical spin control schemes. The latter may result from the lattice response to the charge evolution necessary to perform the spin rotation [7] and from the finite lifetime of the trion excitation used in the optical control schemes. Another source of discrepancies from the desired spin rotation (errors of the single-qubit gate) is imperfect adiabaticity of the evolution. Possible optimization against these errors will be discussed.

## 2. Band structure and selection rules for interband transitions in III-V semiconductors

Quantum dots are semiconductor structures in which the carrier dynamics is restricted in all three dimensions to the length scales of several or a few tens of nanometers [8]. Since the effective mass of carriers is usually considerably lower than the free electron mass, this degree of confinement is sufficient for quantization of carrier energies with electron inter-level spacing reaching 100 meV in self-assembled structures.

The valence band in III-V semiconductors is composed of p-type atomic orbitals (orbital angular momentum 1), yielding six quantum states (taking spin into account) for each quasimomentum $k$. Due to considerable spin-orbit interaction the orbital angular momentum and spin are not separate good quantum numbers and the valence band states of a bulk crystal must be classified by the total angular momentum and its projection on a selected axis. Thus, the valence band is composed of three subbands corresponding to two different representations of the total angular momentum $J$. Out of these, the two states with $j = 1/2$ form a subband which is split-off by the spin-orbit interaction. The other four states with $j = 3/2$ are degenerate in bulk at $k = 0$ but this degeneracy is lifted by size quantization and strain in a QD structure, with the heavy hole (hh) subband (angular momentum projection on the symmetry axis $m = \pm 3/2$) lying above the light hole (lh) subband ($m = \pm 1/2$) in all known struc-tures.

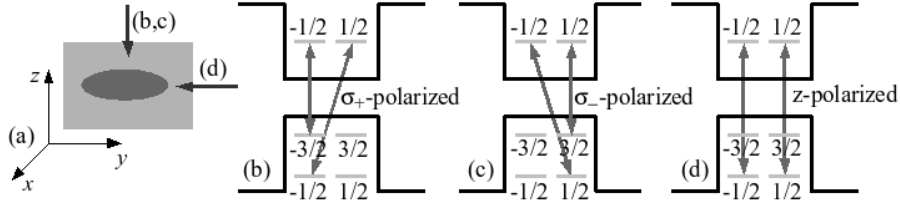

Fig. 1. (a) Schematic plot of a QD. The *xy* plane is referred to as the structure plane, while *z* is called growth direction and corresponds to the symmetry axis of the structure. The arrows show the incidence of light corresponding to the three diagrams b-d. (b-d) The diagrams showing the transitions allowed by selection rules in a III-V QD, induced by light circularly polarized in the structure plane (b: right-polarized, c: left-polarized) and by light linearly polarized along z (d).

The fundamental optical excitation of a QD consists in transferring optically an electron from the highest confined state in the valence band (thus leaving a hole) to the lowest confined state in the conduction band. The interacting electron-hole pair created in this way is referred to as exciton (lh-exciton or hh-exciton, depending on the kind of hole involved). Angular momentum selection rules restrict the transitions allowed for a given propagation direction and polarization of the light beam, as depicted in Fig. 1. For instance, according to the selection rules represented in Fig. 1b, a $\sigma_+$-polarized laser beam can only create an exciton with total momentum +1, referred to as "$\sigma_+$ exciton", in accordance with the angular momentum conservation (removing an electron with the angular momentum $m$ is equivalent to the creation of a hole with the angular momentum $-m$). Similarly, a $\sigma_-$-polarized beam creates only a "$\sigma_-$ exciton" with the angular momentum $-1$.

In appropriately doped structures, QDs in the ground system state may be occupied by electrons. An optical excitation in this case corresponds to a transition between a single electron state and a negative trion state, i.e., the state of two electrons and one hole confined in a QD. From the Pauli exclusion principle it is clear that this transition is possible only if the state which is to be occupied by the photocreated electron is free. Hence, in the situation of Fig. 1b, a heavy hole trion may be created if the dot is initially occupied by a "spin up" (+1/2) electron but not if the electron in the dot is in the "spin down" (-1/2) state. This suppressing of the optical transition depending on the spin of the electron in the QD is referred to as Pauli blocking and has been indeed observed experimentally [9].

## 3. Optical spin control

In this section, we present a brief review of the methods that, according to various theoretical studies, may be used to flip (or, in general, coherently rotate in any desired way) the spin of a single electron confined in a QD.

The existence of the Pauli blocking effect described above means that under specific optical driving conditions the evolution of the charge state, in particular the possibility of a transition to the trion state, may depend on the spin of a confined electron. This fact can be used to flip the spin of a single confined electron residing in a QD [2]. To this end one first applies a pulse of appropriate intensity at normal incidence, tuned to the light hole transition, and linearly polarized. Such a pulse may be decomposed into two circularly polarized components with opposite polarizations. Therefore, according to Fig. 1b,c, it can induce a transition to a light hole negative trion state for both initial spin orientations (Fig. 2a). However, since the transition may involve only the unoccupied electron state (due to Pauli blocking) and thus only the corresponding hole state (due to selection rules) the resulting negative trion states are different, as they contain a hole in different angular momentum states (Fig. 2a). The second step of the procedure is to apply a pulse traveling in the structure plane and polarized linearly in the growth direction. Again, as can be seen from Fig. 1d, such a pulse induces a transition in both cases. In the final state the orientation of the electron spin is always opposite to the initial one (Fig. 2b).

In order to perform an entanglement-generating two-qubit gate on electron spins confined in two dots one can exploit the Coulomb coupling between static electric dipoles formed by the trions, following the procedure proposed for trapped atoms [10,2]. A $\sigma_+$-polarized laser pulse tuned to the heavy hole transition excites the trion only if the electron residing in the dot is in the +1/2 state (otherwise, the transition is Pauli-blocked). By applying such pulses to the two dots one generates the two trions and thus effectively switches on their Coulomb interaction only if both spins are "up". If the trions are present for an appropriate time before they are optically deexcited, the quantum phase accumulated due to the interaction may be equal to $\pi$, thus realizing the standard quantum conditional-phase gate [11], taking the separable state $(|--\rangle + |-+\rangle + |+-\rangle + |++\rangle)/2$ into the entangled state $(|--\rangle + |-+\rangle + |+-\rangle - |++\rangle)/2$.

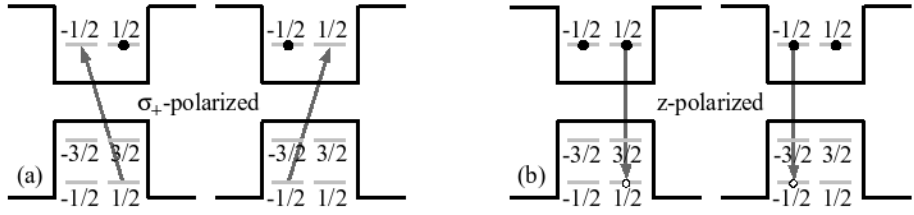

Fig. 2. Spin flip procedure via excitation of light hole trion states. Diagrams (a) and (b) show the two steps of the control procedure. The left diagrams correspond to the initial "spin down" state, and the right diagrams to the "spin up" state.

The procedure presented above shows that purely optical control of spin degrees of freedom is possible, in spite of the absence of a direct spin-light coupling. However, this approach has a few serious disadvantages.

First, the single-qubit rotation is based on the light hole trion transition. The light hole states are energetically higher that the heavy holes which will lead to dephasing due to relaxation. Moreover, coherent optical control of the light hole transition has never been demonstrated in quantum dots. In contrast, the heavy hole transition may be controlled very efficiently, as demonstrated, e.g., by the observation of Rabi oscillations [12].

Second, this scheme involves a non-vanishing occupation of the trion state. This state suffers from radiative decay (on the time scale of 1 ns) which will again decrease the fidelity of the process. Since the two-qubit gate is performed by phase accumulation in the trion state, the latter must be maintained over the time period of $\pi/\Delta E$, where $\Delta E$ is the Coulomb shift of the bi-trion state, which is of the order of 1 meV. Thus, a certain level of error is unavoidable.

Third, it is known that charge evolution in quantum dots is accompanied by phonon excitations, since the phonon modes tend to adapt to their new equilibrium which evolves following the modification of the charge distribution [13,14]. In order to avoid the resulting dephasing the evolution must be carried out slow enough (adiabatically with respect to the phonon modes). As has been discussed for a simple QD charge qubit [15], this adiabaticity requirement together with the finite lifetime of the relevant states puts one into a tradeoff situation, where the error cannot be pushed down below a certain level (of the order of $10^{-3}$).

The optical control procedure described above can be cast into an adiabatic version [3]. By slowly switching on the laser field and smoothly changing its frequency (chirping) one adiabatically takes the initial "spin up" state into a dressed state with a certain admixture of the trion state. According to the adiabatic theorem, after the laser pulse is switched off, the system is brought back to the single electron state but a certain phase is gained during the adiabatic evolution. It can be shown [3] that this phase has an additional contribution if both spins were initially "up" so that, again, the quantum control-phase gate can be performed. Slow evolution eliminates the phonon response (the lattice deformation follows adiabatically). The drawback of this procedure is the need for frequency chirping. Moreover, since the two-qubit gate is based on the phase accumulated from the trion-trion interaction, a considerable trion contribution must be induced and the finite lifetime problem remains.

In order to overcome, at least partly, the potential difficulties incurred by this approach two alternative control schemes have been proposed. In both of them, the qubit states $|0\rangle$ and $|1\rangle$ are defined in terms of spin states in a magnetic field along the $x$ axis, that is, in the structure plane. Since these states can be considered as superpositions of the original +1/2 and -1/2 spin states, both of them can be coupled to a single trion state.

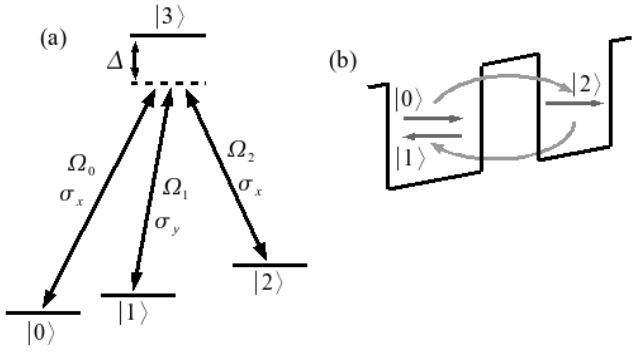

Fig. 3. (a) Schematic representation of states relevant to the spin rotation via Raman adiabatic passage. $\sigma_{x,y}$ denote the polarizations of the pulses. (b) Definition of spin qubit and auxiliary states in terms of spin orientations and spatial location in the double dot structure.

In one of the proposals [5] a double-dot structure is used and the spin rotation is performed via an auxiliary state $|2\rangle$, localized in the second dot. If the hole state forming the trion is delocalized over the two dots then this auxiliary state is also coupled to the trion. Different linear polarizations of the laser beam selectively couple to the two spin states in the dot 1. The auxiliary state may also be addressed selectively if the transition energy in the second dot is different, which is easy to achieve. The diagram of energy levels and laser beams used in this scheme is shown in Fig. 3.

It has been shown within a general framework [16] that such a "three-leg" system allows one to couple nonresonantly (with a detuning $\Delta$ between the transition energy and the laser frequency) a selected (bright) superposition of the qubit states, $|\psi\rangle = \cos\theta\,|0\rangle + e^{-i\varphi}\sin\theta\,|1\rangle$, to the trion state, transfer it via adiabatic Raman passage to the auxiliary state and bring it back with a different phase of the laser pulse. As a result of such a procedure, this specific superposition gains a certain phase $\alpha$ with respect to the orthogonal (dark) one, $|\psi_\perp\rangle = \sin\theta\,|0\rangle - e^{-i\varphi}\cos\theta\,|1\rangle$. It is straightforward to see that the unitary operator corresponding to the resulting transformation can be written as

$$|\psi\rangle\langle\psi|e^{i\alpha} + |\psi_\perp\rangle\langle\psi_\perp| = e^{i\alpha/2}[\cos(\alpha/2)\,I - i\sin(\alpha/2)\,\boldsymbol{n}\cdot\boldsymbol{\sigma}],$$

where I is the unit operator, $\boldsymbol{\sigma}$ is the vector of Pauli matrices, and $\boldsymbol{n} = (\sin 2\theta\cos\varphi, \sin 2\theta\sin\varphi, \cos 2\theta)$. Thus, the procedure results in a rotation of the spin by the angle $\alpha$ around the direction defined by the vector $\boldsymbol{n}$. All the parameters of the rotation can be set by the appropriate choice of relative pulse amplitudes and phases, while the absolute amplitude $\Omega = (\Omega_0^2 + \Omega_1^2 + \Omega_2^2)^{1/2}$ and the detuning $\Delta$ remain arbitrary. An essential feature of this procedure is that for any parameters, the evolution starting in the subspace spanned by $|0\rangle$ and $|1\rangle$ remains in the subspace spanned by the single-electron states $|0\rangle$, $|1\rangle$ and $|2\rangle$ (in a trapped state [17]). Thus, the trion state is never occupied, which eliminates the errors due to its finite lifetime. In Sec. 5 we will see that the procedure may be performed in such a way that phonon decoherence is to a large extent suppressed.

The other proposal [6] is based on the same idea of performing the spin rotation by changing the phase of a selected superposition of the spin state. Now, however, the auxiliary state is absent, which removes the requirement for the very special, delocalized hole state. An arbitrary spin rotation is achieved by an adiabatic evolution within the "lambda system" (so called because of the form of its graphical representation) formed by the two spin states and the trion state. The dynamics of the system may be understood in terms of the AC Stark shift: Since the bright superposition state is coupled to the laser field its energy is shifted. Hence, during the time when the laser pulse is present it accumulates an additional dynamical phase with respect to the dark state. The coupling is again nonresonant and the evolution is adiabatic. The accumulate phase, hence the angle of rotation of the spin, depends on the interplay of the detuning and the absolute pulse intensity, so that one of these parameters may be considered arbitrary and serve for optimization against dephasing.

In general, this approach assumes a nonvanishing occupation of the trion state during the procedure but for large enough detuning this admixture of the trion state is small. Still, the question arises to what extent this small trion admixture may destroy the quantum coherence due to its radiative de-cay. Moreover, as in all the other cases, the dephasing due to phonon response to the charge evolution is an issue. This will be discussed in Sec. 5.

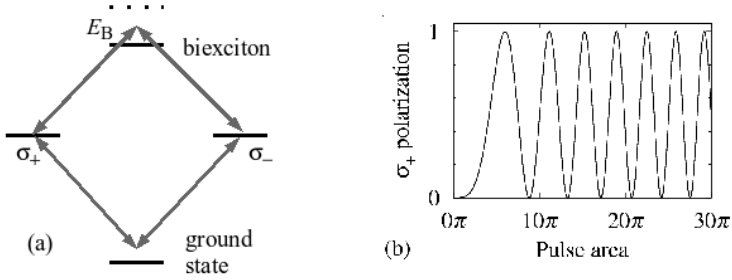

Fig. 4. (a) Schematic representation of the four-level system formed by the empty dot state, two exciton states with different polarizations, and the biexciton state. The arrows show the optical transitions allowed by selection rules for a linearly polarized beam and represent the detunings of the laser beams as applied in the control scheme. (b) The occupation of the $\sigma_+$-polarized exciton state after illumination by a linearly polarized laser pulse as a function of the pulse area. The system has been initially prepared in the $\sigma_-$-polarized exciton state.

Apart from optical control schemes designed for free-standing dots, there are also proposals focused on QDs placed in microcavities. Single-spin rotations are then performed by Raman transitions induced by an external laser beam, exploiting the coupling to the trion state. This can be done either by using light hole states [4] or by defining the spin qubit states with respect to the *x* axis in a magnetic field along this axis [1]. The advantage of the cavity setup is that strong coupling to the cavity mode may provide an effective long-range coupling between spins. In this way, two-qubit gates may be performed not only on neighboring spins as it was in the dipole interaction-based schemes discussed earlier. It allows also for larger separations between the dots so that they can be addressed individually by laser beams which is not possible if the dots must be spaced close enough (on distances at most of the order of 10 nm) to generate a reasonable Coulomb interaction [18,19].

## 4. Optical reorientation of the exciton polarization

There is another possibility related to the optical control of carrier spins in QDs. As mentioned in Sec. 2, apart from single electrons originating from doping, quantum dots can confine electron-hole pairs, referred to as excitons. Such an excitation may be created optically by transferring an electron form the highest confined state in the valence band (thus leaving a hole) to the lowest confined state in the conduction band. We will restrict the discussion to the heavy hole subband. From the diagrams in Fig. 1 it is clear that the excitons of different kinds (polarizations $\sigma_+,\sigma_-$) are composed of particles with different spins, so that no Pauli blocking is present and they can coexist, forming a four-particle complex called a biexciton. Due to the Coulomb interaction between the particles, the energy of the biexciton is shifted with respect to twice the exciton energy (the magnitude $E_B$ of this shift is referred to as the biexciton binding energy, even though it is not always negative). The resulting diagram of quantum states (restricted to the lowest orbital state) is presented in Fig. 4a.

In this section, we will discuss a method for simultaneously changing the spins of both the electron and the hole forming a confined exciton. Thus, the spin orientation (polarization) of an exciton confined in a QD may be controlled with a laser field.

According to the selection rules, a linearly polarized laser beam couples the exciton states with both polarizations to the ground and biexciton states. Since the biexciton state is shifted by $E_B$, as discussed above, the beam may be detuned by $\Delta = E_B/2$ from all the transitions (see Fig. 4a). Thus, a two-photon transition between the two bright exciton states is possible. On the other hand, the detuning of the laser pulses guarantees that the transition to the ground or biexciton state is forbidden by energy conservation.

A theoretical analysis shows that the system follows an adiabatic evolution, resulting in a polarization rotation due to the phase accumulation, similar to that discussed in the previous section. This process is fully coherent (neglecting phonon effects and radiative decay of the states involved) and Rabi oscillations of the exciton polarization can be induced under suitable driving conditions. The latter have a very specific pulse-area dependence, characteristic of two-photon processes (Fig. 4b).

In fact, this two-photon process is strictly analogous to the two-photon Rabi oscillations of a biexciton system demonstrated in a recent experiment [19] and is expected to take place under the same conditions. The polarization of an exciton is relatively easily measurable via the polarization of the photons emitted in the radiative decay process.

**5. Phonon dephasing in the optical spin control schemes**

So far, we have discussed the optical spin control schemes in the idealized situation, as if the QDs were isolated systems, interacting only with the driving fields. However, a quantum dot is always embedded in a solid state matrix and the confined carriers interact with phonons, which leads to phonon excitations accompanying the charge evolution. As the spin rotation in the optical schemes results from charge dynamics, it is subject to dephasing due to the phonon response to the evolving charge density [7].

In this section, the interplay of various kinds of decoherence mechanisms is studied for an optical control of the spin in a doped semiconductor quantum dot within the last two approaches discussed in Sec. 3 [5,6]. We will consider a single-qubit gate: an arbitrary rotation between two selected electron spin states.

Let us start with the scheme employing the auxiliary state [5]. Since in this scheme the single, uncompensated charge carrier is shifted between different spatial locations, one may expect a considerable phonon response due to both deformation potential and piezoelectric coupling (in piezoelectric compounds, like GaAs). Therefore, in the presence of phonons, the fundamental adiabaticity condition, required to perform the quantum-optical procedure of Raman transfer, must be supplemented by the additional requirement to avoid phonon-assisted processes.

A formal analysis [20] has shown that the carrier-phonon coupling indeed gives rise to pure dephasing as well as to transitions between the trapped carrier-field states. The former decreases for slower driving [15], while the contribution from the latter is approximately proportional to the process duration. The probability of phonon-induced transitions becomes very high if the spacing between the trapped energy levels falls into the area of high phonon spectral density. For a double-dot structure, the latter has a complicated, oscillatory form. Moreover, in an attempt to optimize the control parameters one encounters a trade-off situation, due to the opposite requirements for phonon-induced jumps (short duration) and for the fundamental adiabaticity condition and pure dephasing (slow operation). Therefore, in order to estimate the fidelity of the spin rotation and to find optimal parameters for the control procedure one has to perform detailed quantitative calculations taking into account the details of the system structure.

As was mentioned in Sec. 3, this control procedure allows for a free choice of the absolute beam intensity $\Omega$ and of the detuning $\Delta$. Together with the arbitrary duration of the control pulses, this yields three free parameters. The phonon-induced decoherence effects may be determined using the general perturbative approach that allows one to calculate the phonon perturbation for an arbitrary driven evolution of the system [7]. In Fig. 5a we plot the error $\delta = 1-F^2$, where $F$ is the fidelity of the operation (a standard measure of the overlap between the desired final state and the actual one [11]), minimized over the pulse durations, as a function of the other two parameters. It is clear that this procedure may be performed with errors very low compared to the threshold value of $10^{-4}$ or $10^{-3}$ [21] required for the implementation of quantum error-correction schemes and thus for a scalable implementation of quantum computing. However, such a high fidelity is achievable only in very narrow ranges of parameter values.

In the control scheme without the auxiliary state [6], there are again contributions to the error from the dynamical phonon response (pure dephasing), from jumps between the adiabatic states, and from the imperfect adiabaticity of the evolution. The jumps may now be interpreted as phonon-assisted trion generation and their impact is particularly large for positive detuning, when a transition to the trion state may take place upon absorption of a photon and emission of a phonon, which is possible even at zero temperature. Another contribution to decoherence now originates from the finite lifetime of the trion since, in contrast to the previous control scheme, a certain occupation of the trion state appears during the evolution.

In order to assess the importance of all these contributions one has to perform detailed calcula-tions, including the carrier-phonon couplings, radiative decay of the trion state and corrections from the nonadiabaticity of the evolution. Since the angle of rotation of the spin now depends on a combination of the absolute pulse intensity and detuning only one of these parameters may be considered free. Another free parameter is the pulse duration. In Fig. 5b we plot the error for a π/2 rotation, averaged over the initial state, as a function of the detuning $\Delta$ (pulse amplitudes are then adjusted to achieve the π/2 rotation) for a few values of the pulse duration. A positive detuning means a red-shift of the laser frequencies with respect to the trion transition. As a result of the complicated interplay of the different contributions, the error now depends on the detuning in a quite irregular way. It turns out that the fidelity of the operation strongly depends on the pulse duration only for sub-meV detuning, where the error is relatively high. Very high fidelity may be achieved for a detuning of the order of 10 meV.

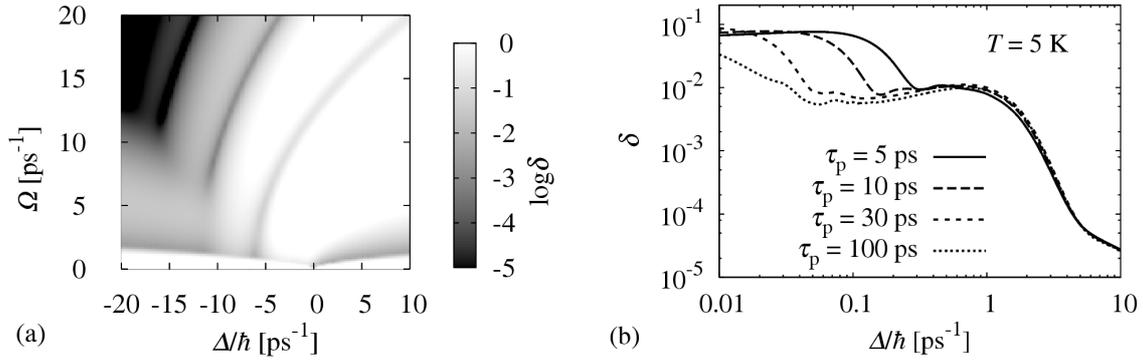

Fig. 5. (a) The error for the spin rotation performed using the Raman adiabatic passage, as a function of the laser pulse parameters, for the pulse durations of 50 ps at 1 K. (b) The error for a π/2 spin rotation without the auxiliary state as a function of the detuning for a few pulse durations. Both results correspond to a self-assembled InAs/GaAs structure.

## 6. Conclusion

We have reviewed a few theoretical proposals showing that optical control of individual spins and exciton polarizations in quantum dots is possible, at least in principle. We have shown that careful optimization of parameters may bring the errors of such optical spin rotations (including those related to carrier-phonon interactions) to very low levels. Some of these proposals are very demanding with respect to the system properties and, so far, none of them has been realized experimentally. However, the rapid progress of the engineering of QD structures and coherent optical control of quantum dots that can be observed in recent years will certainly make the implementation of such control techniques feasible in the close future.

**References**
1. İmamoğlu A., Awschalom D. D., Burkard G., DiVincenzo D. P., Loss D., Sherwin M. and Small A., Phys. Rev. Lett., 83 (1999), 4202.
2. Pazy E., Biolatti E., Calarco T., D'Amico I., Zanardi P., Rossi F. and Zoller P., Europhys. Lett., 62 (2003), 175.
3. Calarco T., Datta A., Fedichev P., Pazy E. and Zoller P., Phys. Rev. A, 68 (2003), 012310.
4. Feng M., D'Amico I., Zanardi P. and Rossi F., Phys. Rev. A, 67 (2003), 014306.
5. Troiani F., Molinari E. and Hohenester U., Phys. Rev. Lett., 90 (2003), 206802.
6. Chen P., Piermarocchi C., Sham L. J., Gammon D. and Steel D. G., Phys. Rev. B, 69 (2004) 075320.
7. Grodecka A., Jacak L., Machnikowski P., Roszak K., in: *Quantum Dots: Research Developments*, P. A. Ling (Ed), Nova Science, New York, 2005, p. 47; cond-mat/0404364.
8. Jacak L., Hawrylak P. and Wójs A., Quantum Dots, Springer, Berlin, 1998.
9. Chen G., Bonadeo N. H., Steel D. G., Gammon D., Katzer D. S., Park D. and Sham L. J., Science, 289 (2000), 1906.
10. Jaksch D., Cirac J. I., Zoller, P., Rolston S. L., Côté R. and Lukin M. D., Phys. Rev. Lett., 85

(2000), 2208.
11. Nielsen M. A. and Chuang I. L., Quantum Computation and Quantum Information, Cambridge University Press, Cambridge, 2000.
12. Zrenner A., Beham E., Stuffer S., Findeis F., Bichler M. and Abstreiter G., Nature, 418 (2002), 612.
13. Jacak L., Machnikowski P., Krasnyj J. and Zoller P., Eur. Phys. J. D, 22 (2003), 319.
14. Vagov A., Axt V. M. and Kuhn T., Phys. Rev. B, 66 (2002), 165312.
15. Alicki R., Horodecki M., Horodecki P., Horodecki R., Jacak L. and Machnikowski P., Phys. Rev. A, 70 (2004), 010501(R).
16. Kis Z. and Renzoni F., Phys. Rev. A, 65 (2002), 032318.
17. Scully M. O. and Zubairy M. S., Quantum Optics, Cambridge University Press, Cambridge, 1997.
18. Danckwerts J., Ahn K. J., Förstner J. and Knorr A., Phys. Rev. B 73 (2006), 165318.
19. Richter M., Ahn K. J., Knorr A., Schliwa A., Bimberg D., Madjet M. E. and Renger T., Phys. Stat. Sol. (b) 243 (2006), 2302.
20. Stufler S., Machnikowski P., Ester P., Bichler M., Axt V. M., Kuhn T. and Zrenner A., Phys. Rev. B, 73 (2006), 125304.
21. Roszak K., Grodecka A., Machnikowski P. and Kuhn T., Phys. Rev. B, 71 (2005), 195333.
22. Knill E., Nature, 434 (2005), 39.